\documentclass[aps,prl,twocolumn,superscriptaddress,showpacs,preprintnumbers,amsmath,amssymb]{revtex4}
\usepackage{graphicx} 
\usepackage{dcolumn}  

\graphicspath{{ps}}

\begin{document}


\preprint{\vbox{ \hbox{}
                 \hbox{}
}}

\title{ \quad\\[0.5cm]  Measurements of Branching Fraction and
  Polarization in $B^+\to \rho^+ K^{*0}$ Decay}


\affiliation{Budker Institute of Nuclear Physics, Novosibirsk}
\affiliation{Chiba University, Chiba}
\affiliation{Chonnam National University, Kwangju}
\affiliation{University of Cincinnati, Cincinnati, Ohio 45221}
\affiliation{Gyeongsang National University, Chinju}
\affiliation{University of Hawaii, Honolulu, Hawaii 96822}
\affiliation{High Energy Accelerator Research Organization (KEK), Tsukuba}
\affiliation{Hiroshima Institute of Technology, Hiroshima}
\affiliation{Institute of High Energy Physics, Chinese Academy of Sciences, Beijing}
\affiliation{Institute of High Energy Physics, Vienna}
\affiliation{Institute for Theoretical and Experimental Physics, Moscow}
\affiliation{J. Stefan Institute, Ljubljana}
\affiliation{Kanagawa University, Yokohama}
\affiliation{Korea University, Seoul}
\affiliation{Kyungpook National University, Taegu}
\affiliation{Swiss Federal Institute of Technology of Lausanne, EPFL, Lausanne}
\affiliation{University of Ljubljana, Ljubljana}
\affiliation{University of Maribor, Maribor}
\affiliation{University of Melbourne, Victoria}
\affiliation{Nagoya University, Nagoya}
\affiliation{Nara Women's University, Nara}
\affiliation{National Central University, Chung-li}
\affiliation{National United University, Miao Li}
\affiliation{Department of Physics, National Taiwan University, Taipei}
\affiliation{H. Niewodniczanski Institute of Nuclear Physics, Krakow}
\affiliation{Nippon Dental University, Niigata}
\affiliation{Niigata University, Niigata}
\affiliation{Nova Gorica Polytechnic, Nova Gorica}
\affiliation{Osaka City University, Osaka}
\affiliation{Osaka University, Osaka}
\affiliation{Peking University, Beijing}
\affiliation{Princeton University, Princeton, New Jersey 08544}
\affiliation{Saga University, Saga}
\affiliation{University of Science and Technology of China, Hefei}
\affiliation{Seoul National University, Seoul}
\affiliation{Sungkyunkwan University, Suwon}
\affiliation{University of Sydney, Sydney NSW}
\affiliation{Tata Institute of Fundamental Research, Bombay}
\affiliation{Toho University, Funabashi}
\affiliation{Tohoku Gakuin University, Tagajo}
\affiliation{Tohoku University, Sendai}
\affiliation{Department of Physics, University of Tokyo, Tokyo}
\affiliation{Tokyo Institute of Technology, Tokyo}
\affiliation{Tokyo Metropolitan University, Tokyo}
\affiliation{Tokyo University of Agriculture and Technology, Tokyo}
\affiliation{University of Tsukuba, Tsukuba}
\affiliation{Virginia Polytechnic Institute and State University, Blacksburg, Virginia 24061}
\affiliation{Yonsei University, Seoul}
  \author{J.~Zhang}\affiliation{High Energy Accelerator Research Organization (KEK), Tsukuba} 
   \author{K.~Abe}\affiliation{High Energy Accelerator Research Organization (KEK), Tsukuba} 
   \author{I.~Adachi}\affiliation{High Energy Accelerator Research Organization (KEK), Tsukuba} 
   \author{H.~Aihara}\affiliation{Department of Physics, University of Tokyo, Tokyo} 
   \author{K.~Arinstein}\affiliation{Budker Institute of Nuclear Physics, Novosibirsk} 
   \author{Y.~Asano}\affiliation{University of Tsukuba, Tsukuba} 
   \author{T.~Aushev}\affiliation{Institute for Theoretical and Experimental Physics, Moscow} 
   \author{T.~Aziz}\affiliation{Tata Institute of Fundamental Research, Bombay} 
   \author{S.~Bahinipati}\affiliation{University of Cincinnati, Cincinnati, Ohio 45221} 
   \author{A.~M.~Bakich}\affiliation{University of Sydney, Sydney NSW} 
% ...
\collaboration{The Belle Collaboration}

\noaffiliation

\begin{abstract}
  We present the results of a study of the charmless vector-vector
  decay $B^+\to \rho^+K^{*0}$,
  based on 253 fb$^{-1}$  of data collected with
  the Belle  detector at the KEKB asymmetric-energy $e^+e^-$ collider.
  We obtain the branching fraction 
  ${\cal B}(B^+\to\rho^+ K^{*0})=(8.9\pm 1.7(\rm
  stat)\pm 1.2 (\rm syst))\times 10^{-6}$.
  We also perform a helicity analysis of the $\rho$ and $K^*$ vector
  mesons, and obtain the longitudinal polarization fraction
  $f_L(B^+\to \rho^+ K^{*0})=0.43 \pm 0.11(\rm
  stat)^{+0.05}_{-0.02}(\rm syst)$.
  
\end{abstract}

\pacs{13.25.Hw, 14.40.Nd.}

\maketitle

\tighten

{\renewcommand{\thefootnote}{\fnsymbol{footnote}}}
\setcounter{footnote}{0}

Naive factorization in the Standard Model (SM) predicts that the
longitudinal polarization fraction ($f_L$) in $B$ meson decays to
light vector-vector (VV) final states is close to
unity~\cite{ref:Kagan}. 
In the tree dominated $B^+ \to \rho^+\rho^0$ and 
$B^0 \to \rho^+\rho^-$ decays, this prediction
has been confirmed~\cite{ref:rhoprho0, ref:BaBar_vv,
  ref:BaBar_rhoprhom}.
In contrast, for the pure $b\to s$ penguin $B \to \phi K^*$ decay,
Belle~\cite{ref:phiK} and BaBar~\cite{ref:BaBar_vv} have found that
the longitudinal and transverse polarization fractions are comparable,
which is in disagreement with the factorization expectation.
Possible explanations for this discrepancy include enhanced
non-factorizable contributions such as penguin
annihilation~\cite{ref:Kagan}, large
$SU$(3) breaking in form factors~\cite{ref:Li}, or new
physics~\cite{ref:Grossman, ref:Yang}.
It is therefore important to perform polarization measurements in other
VV modes, in particular, in the pure penguin $b\to s \overline d d$ decay
$B^+\to \rho^+ K^{*0}$.

In this paper, we present the results of a study of $B^+\to
\rho^+K^{*0}$ decays~\cite{charge_conjugate} with a 253 fb$^{-1}$ data
sample containing $275\times 10^{6}$ $B$ meson pairs collected with
the Belle detector at the KEKB asymmetric-energy $e^+e^-$
collider~\cite{KEKB} operating at the $\Upsilon(4S)$ resonance
($\sqrt{s} = 10.58$~GeV). The production rates for $B^+B^-$ and
$B^0\overline{B}{}^0$ pairs are assumed to be equal.\\

The Belle detector is a large solid-angle magnetic spectrometer that
consists of a silicon vertex detector (SVD),
a 50-layer central drift chamber (CDC), an array of
aerogel threshold \v{C}erenkov counters (ACC), 
a barrel-like arrangement of time-of-flight
scintillation counters (TOF), and an electromagnetic calorimeter
comprised of CsI(Tl) crystals (ECL) located inside 
a superconducting solenoid coil that provides a 1.5~T
magnetic field.  An iron flux-return located outside of
the coil is instrumented to detect $K_L^0$ mesons and to identify
muons (KLM).  The detector is described in detail
elsewhere~\cite{Belle}.\\

We select $B^+ \to \rho^+ K^{*0}$ candidate events by combining
three charged tracks (two oppositely charged pions and one kaon) and
one neutral pion.
Each charged track is required to have a transverse momentum $p_T>0.1$
GeV$/c$ and to have an origin within $0.2~{\rm cm}$ in
the radial direction and $5~\rm{cm}$ along the beam direction of the
interaction point (IP).

Particle identification likelihoods for the pion and kaon
hypotheses are calculated by combining information from the TOF and ACC
systems with $dE/dx$ measurements in the CDC. To identify kaons, we
require the kaon likelihood ratio, $L_K/(L_K+L_\pi)$, to be greater
than 0.6. To identify pions, we require $L_K/(L_K+L_\pi)$ to be less
than 0.4. 
The efficiency for this selection is $86\%$ for kaons and $89\%$ for
pions, with corresponding $\pi/K$ misidentification rates of $8\%$ and
$10\%$.
In addition, charged tracks are rejected if they are consistent with
an electron hypothesis.

Candidate $\pi^0$ mesons are reconstructed from pairs of photons that
have an invariant mass in the range $0.1178$ -- 0.1502 GeV/$c^2$,
corresponding to a window of $\pm 3\,\sigma$ around the nominal
$\pi^0$ mass. The photons are assumed to originate from the IP.
The energy of each photon in the laboratory frame is required to be
greater than 50 MeV for the ECL barrel region ($32^\circ< \theta
<129^\circ$) and 100 MeV for the ECL endcap regions
($17^\circ<\theta<32^\circ$ and $129^\circ<\theta<150^\circ$), where
$\theta$ denotes the polar angle of the photon with respect to the
beam line. The $\pi^0$ candidates are kinematically constrained to the
nominal $\pi^0$ mass. In order to reduce the combinatorial background,
we only accept $\pi^0$ candidates with momenta $p_{\pi^0}>0.40$
GeV$/c$ in the $e^+e^-$ center-of-mass (CM) system.

Candidate $\rho^+$ mesons are reconstructed via their $\rho^+ \to \pi^+
\pi^0$ decay, and the $\pi^+\pi^0$ pairs are required to have an
invariant mass in the region $0.62{~\rm
  GeV}/c^2<M(\pi^+\pi^0)<0.92{~\rm GeV}/c^2$.
Candidate $K^{*0}$ mesons are selected from the $K^{*0}\to
K^+\pi^-$ decay with an invariant mass $0.83{~\rm
  GeV}/c^2<M(K^+\pi^-)<0.97{~\rm GeV}/c^2$.

To isolate the signal, we form the beam-constrained mass $M_{\rm
  bc}\equiv\sqrt{E_{\rm beam}^2-p_B^2}$, and the energy difference
$\Delta E\equiv E_B-E_{\rm beam}$, where $E_{\rm beam}$ is the CM
beam energy, and $p_B$ and $E_B$ are the CM momentum and energy,
respectively, of the $B$ candidate.
The $\Delta E$ distribution has a tail on the lower side caused by
incomplete longitudinal containment of electromagnetic showers in 
the CsI(Tl) crystals.
We accept events in the region 
$M_{\rm bc}>5.2 {~\rm GeV}/c^2$ and $-0.3{~\rm GeV}<\Delta E<0.3{~\rm GeV}$, 
and define a signal region in $M_{\rm bc}$ and $\Delta E$ as 
$5.27{~\rm GeV}/c^2<M_{\rm bc}<5.29{~\rm GeV}/c^2$ and
$-0.10{~\rm GeV}<\Delta E<0.06{~\rm GeV}$ respectively.
These requirements correspond to approximately $\pm 3\,\sigma$ for both
quantities.

The continuum process $e^+e^- \to q \overline q$ ($q=u, d, s, c$) is the
main source of background and must be strongly suppressed. 
One method of discriminating the signal from continuum is
based on the event topology, which tends to be isotropic for $B\overline B$
events and jet-like for $q\overline q$ events.  
Another discriminating characteristic is $\theta_B$, the CM polar
angle of the $B$ flight direction.
$B$ mesons are produced with a $1-\cos^2\theta_B$ distribution
while continuum background events tend to be uniform in
$\cos\theta_B$.
The displacement along the beam direction between the signal $B$
vertex and that of the other $B$, $\Delta z$, also provides
separation.
For $B$ events, the average value of $\Delta z$ is approximately 200
$\mu$m, while continuum events have a common vertex.
Additional discrimination is provided by the $b$-flavor tagging
algorithm~\cite{ref:kakuno} developed for time-dependent analysis at
Belle.
The flavor tagging procedure yields two outputs: 
$q$ $(=\pm 1)$, which indicate the flavor of the tagging $B$,
and $r$, which ranges from 0 to 1, is a measure of the likelihood that
the $b$ flavor of the accompanying $B$ meson is correctly assigned.
For signal events, $q$ is more likely consistent with the opposite of
the charge of signal $B$; there is no correlation for continuum
events.
Events with high values of $r$ are well-tagged and are less likely
to originate from continuum production.
Thus, the quantity $q \cdot r \cdot C_B$, where $C_B$ is the charge of
the signal $B$, can be used to discriminate against continuum events.

We use Monte Carlo (MC) simulated signal and data sideband (defined as
$5.2~{\rm GeV}/c^2<M_{\rm bc}<5.26~{\rm GeV}/c^2$) events to form a
Fisher discriminant based on a set of modified Fox-Wolfram
moments~\cite{shlee} that are confirmed to be uncorrelated with $M_{\rm
  bc}$, $\Delta E$ and variables considered later in the analysis.
Probability density functions (PDFs) derived from the Fisher
discriminant, the $\cos\theta_B$ distributions and the $\Delta z$
distributions are multiplied to form likelihood functions for signal
(${\cal L}_{s}$) and continuum background (${\cal L}_{q\overline q}$); 
these are combined into a likelihood ratio ${\cal R}_s={\cal
  L}_{s}/({\cal L}_{s}+{\cal L}_{q\overline q})$.
We achieve background suppression by imposing $q\cdot r\cdot
C_B$-dependent ${\cal R}_s$ requirements, which are determined by
optimizing the figure of merit, $S/\sqrt{S+B}$, where $S$ ($B$) is
the number of signal (background) events  in the signal region.
A branching fraction of ${\cal B}(B^+\to \rho^+K^{*0})=1\times
10^{-5}$ is assumed.
This requirement removes 99.3\% of the continuum background while
retaining 41\% of the $B^+ \to \rho^+ K^{*0}$ events.
The MC-determined efficiency with all selection criteria imposed is
2.7\% for longitudinal polarization ($A_0$) and 4.0\% for transverse
polarization ($A_\pm$).

The fraction of multiple candidates in the signal region for signal MC
is 3.6\% for the $A_0$ helicity state and 1.7\% for the $A_\pm$
state. We allow multiple candidates in this analysis.

To investigate backgrounds from $b\to c$ decays, we use a
sample of $B\overline B$ MC events corresponding to an integrated
luminosity of 412 fb$^{-1}$. 
We find a contribution from $B^+ \to \overline D{}^0(K^+\pi^-
\pi^0)\pi^+$ decays in the $\rho$ or $K^*$ sideband region
and require $|M(K \pi \pi^0)-M_{D^0}|>0.050$ GeV$/c^2$ to veto these
events.
This requirement does not remove any $B^+ \to \rho^+ K^{*0}$ events. 
Among the charmless $B$ decays, potential backgrounds arise from
$B^+\to a_1^0 K^+$, $B^+ \to \rho^+ K_0^*(1430)$, non-resonant $B^+\to
\rho^+ K^+\pi^-$ and $B^+ \to K^{*0} \pi^+ \pi^0$.
We separate signal from these backgrounds by fitting the $\rho$ and
$K^*$ invariant mass distributions.

\begin{figure}[htbp]
  \begin{center}
    \includegraphics[width=4.2cm,clip]{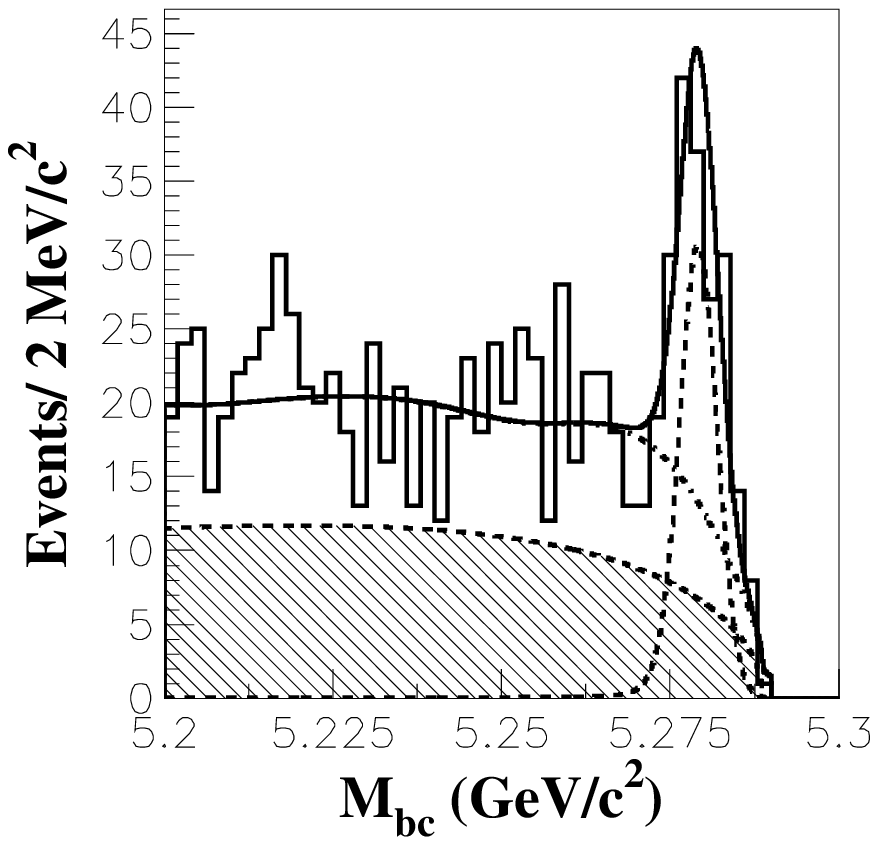}
    \includegraphics[width=4.2cm,clip]{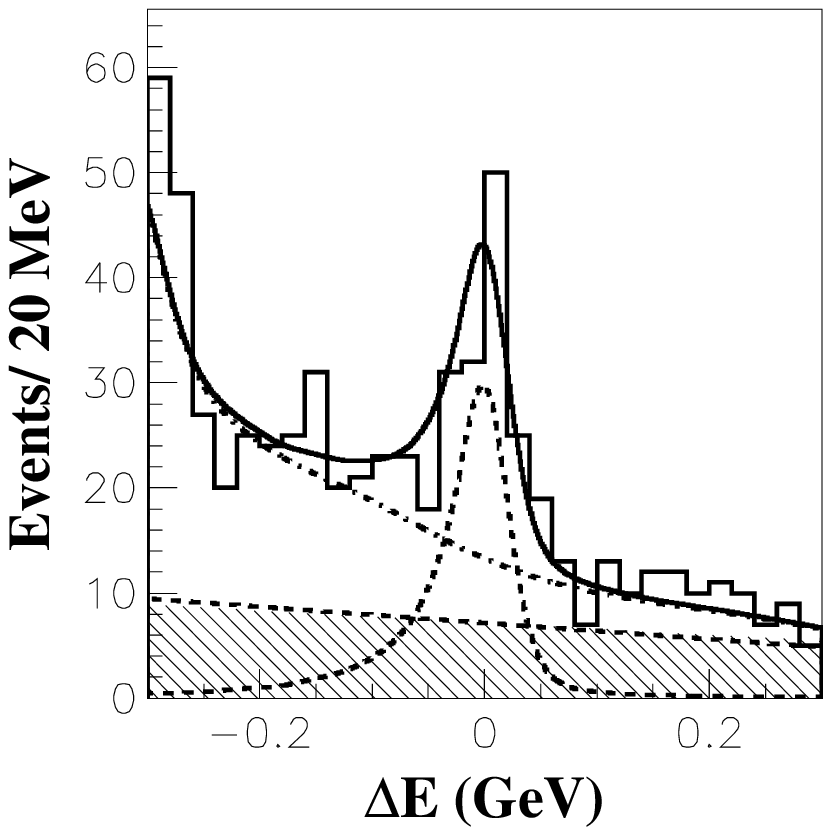}
    \caption{Projections of $M_{\rm bc}$ for events in the $\Delta E$
      signal region (left), and projection of $\Delta E$ in the $M_{\rm
	bc}$ signal region (right). 
      The solid curves show the results of the fit. The hatched
      histograms represent the continuum background. The sum of the $b\to
      c$ and continuum background component is shown as dot-dashed lines.}
    \label{fit_mbde}
  \end{center}
\end{figure}

We extract the signal yield by applying an extended unbinned
maximum-likelihood fit to the two-dimensional $M_{\rm bc}$-$\Delta E$
distribution. The fit includes components for signal plus backgrounds
from continuum events and $b\to c$ decays.
The PDFs for signal and $b\to c$ decay are modeled by smoothed
two-dimensional histograms obtained from large MC samples.
The signal PDF is adjusted to account for small differences
observed between data and MC for a high-statistics mode
containing $\pi^0$ mesons, $B^+ \to {\overline
  D}{}^0(K^+\pi^-\pi^0)\pi^+$.
The continuum PDF is described by a product of a threshold (ARGUS)
function~\cite{ref:ARGUS} for $M_{\rm bc}$ and a first-order polynomial
for $\Delta E$, with shape parameters allowed to vary.
All normalizations are allowed to float.
Figure \ref{fit_mbde} shows the final event sample and the fit results.
The five-parameter (three normalizations plus two shape parameters for
continuum) fit yields $134.8\pm 16.9$ $B^+\to K^+\pi^-\pi^+\pi^0$
events.

\begin{figure}[htbp]
  \begin{center}
    \includegraphics[width=4.2cm,clip]{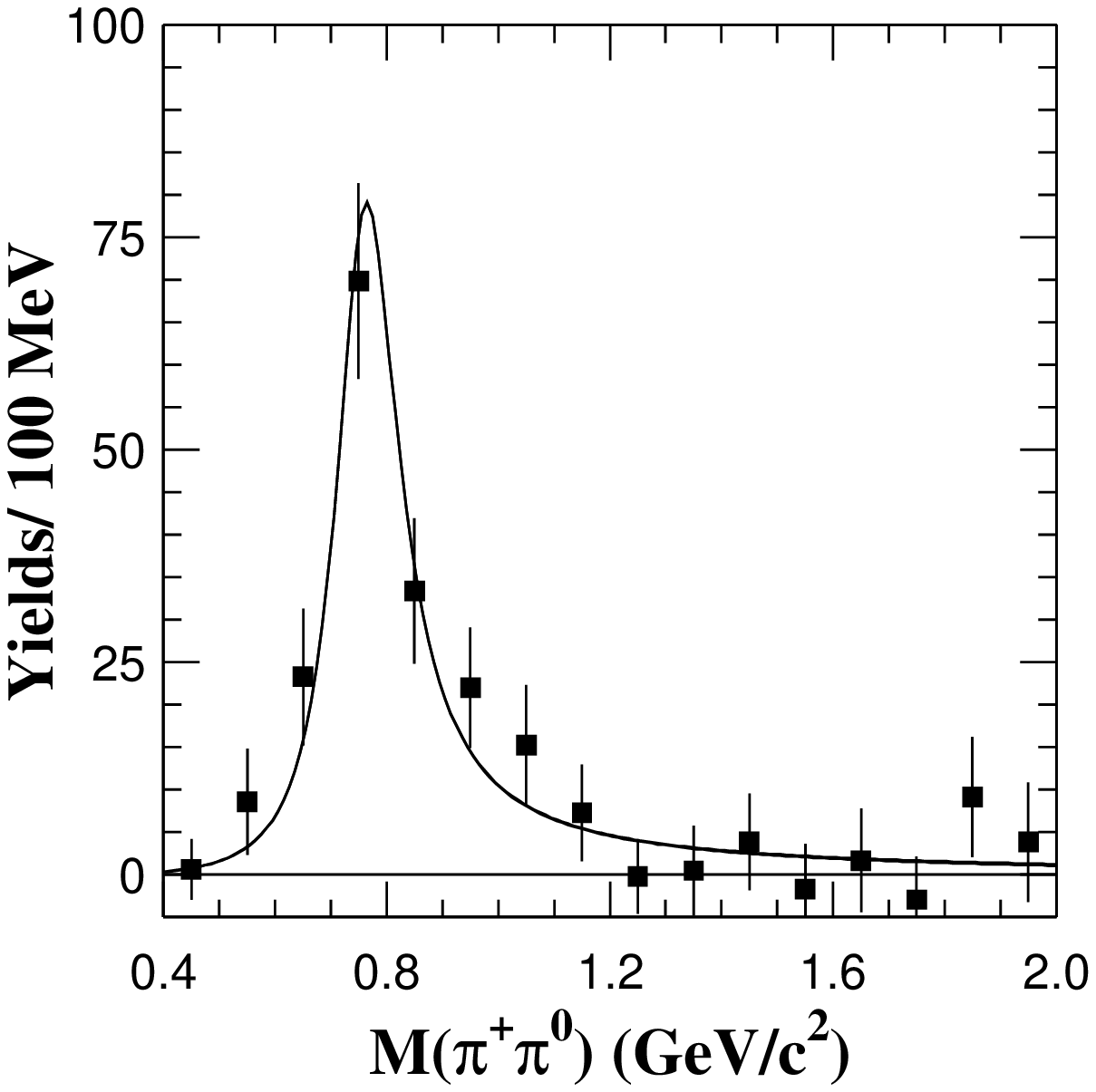}
    \includegraphics[width=4.2cm,clip]{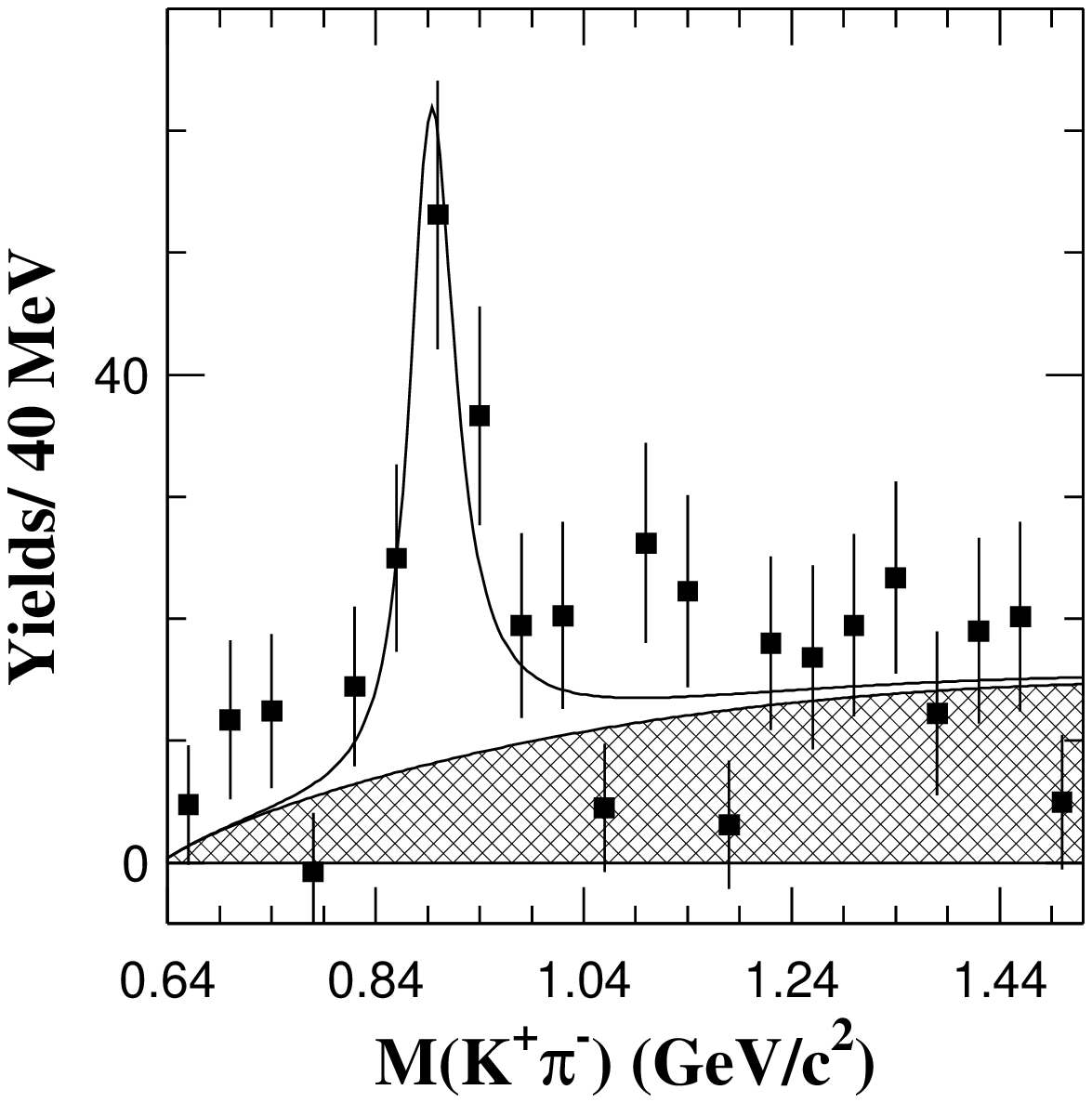}
    \caption{Signal yields obtained from the $M_{\rm bc}$-$\Delta E$
      distribution in bins of $M(\pi^+\pi^0)$ (left) for events in
      the $K^{*0}$ region and in bins of $M(K^+\pi^-)$ (right) for
      events in the $\rho$ region.
      Solid curves show the results of the fit.
      Hatched histograms are for the non-resonant component.}
    \label{fit_mass}
  \end{center}
\end{figure}

We further distinguish the $\rho^+ K^{*0}$ signal from non-resonant
decays such as $B^+ \to \rho^+ K^+\pi^-$ or $B^+\to
K^{*0}\pi^+\pi^0$ by fitting the $M(\pi^+\pi^0)$ and $M(K^+\pi^-)$
invariant mass distributions.
The signal yields obtained from the $M_{\rm bc}$-$\Delta E$ fit for
different $M(\pi\pi)$ and $M(K\pi)$ bins are plotted in
Fig.~\ref{fit_mass},
where the $M(\pi\pi)$ distribution is for events in the $K^*$
region ($0.83{~\rm GeV}/c^2 <M(K\pi)<0.97{~\rm GeV}/c^2$) and the
$M(K\pi)$ distribution is for events in the $\rho$ region ($0.62{~\rm
  GeV}/c^2<M(\pi\pi)<0.92$ GeV/$c^2$).
We perform separate $\chi^2$ fits to the 
$M(\pi\pi)$ or $M(K\pi)$ distributions.
Each fit includes components for signal and non-resonant background.
The signal $\rho$ and $K^*$ PDFs are modeled by relativistic $P$-wave
Breit-Wigner functions with means and widths fixed at their known
values~\cite{PDG};
the PDFs are convolved with a Gaussian of $\sigma=5.3$ MeV, which is
obtained by fitting the $D^0(K^-\pi^+)$ invariant mass, to account for
the detector resolution.
The non-resonant component is represented by a threshold function with
parameters determined from MC events where the final states are
distributed uniformly over phase space.
The $M(\pi\pi)$ mass fit gives $125.4\pm15.8$ $\rho$ and $-0.3\pm3.0$
non-resonant  $K^{*0}\pi^+\pi^0$ events in the $\rho$ mass region.
In the $M(K\pi)$ fit, we find $85.4\pm16.1$ $\rho K^*$ signal and  
$28.8\pm 4.1$ non-resonant events in the $K^*$ mass region. 
The statistical significance of the signal, defined as
$\sqrt{\chi_0^2-\chi_{\rm min}^2}$, where $\chi_{\rm
  min}^2$ is the $\chi^2$ value at the best-fit signal  yield and
$\chi_0^2$ is the value with the $K^{*0}$ signal yield set to
zero, is $5.3\,\sigma$ ($5.2\,\sigma$ with the inclusion of
systematics).
The contribution from non-resonant $\rho^+ K^+\pi^-$ is significant
and is taken into account in both the branching fraction and
polarization determinations, while we neglect the non-resonant
$K^{*0}\pi^+\pi^0$ contribution. 

\begin{figure}[htbp]
  \begin{center}
    \includegraphics[width=4.2cm,clip]{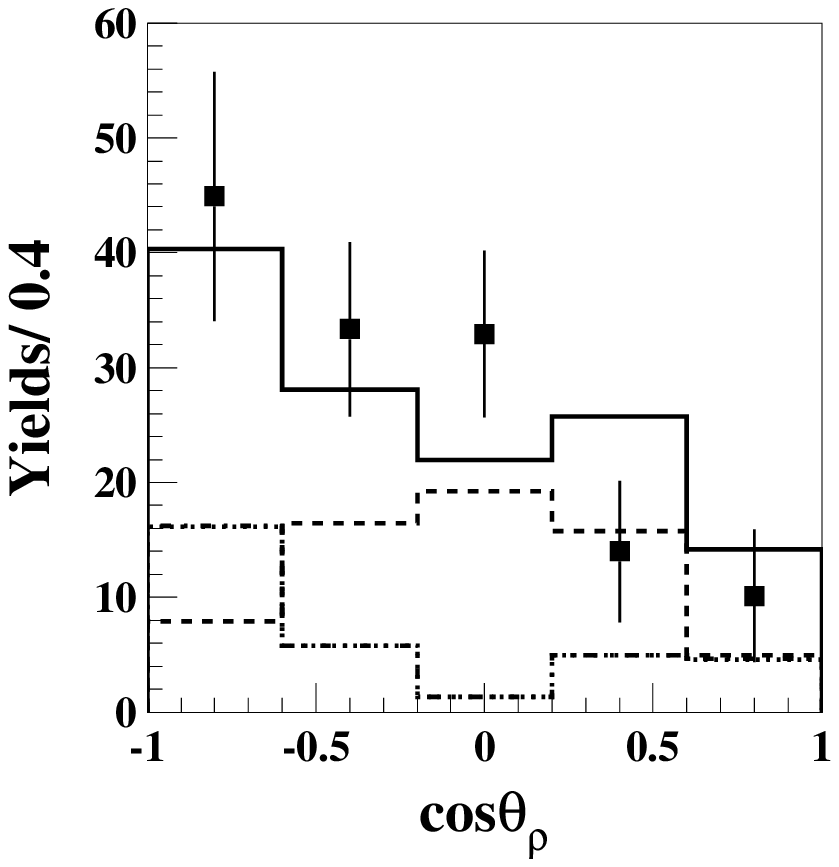}
    \includegraphics[width=4.2cm,clip]{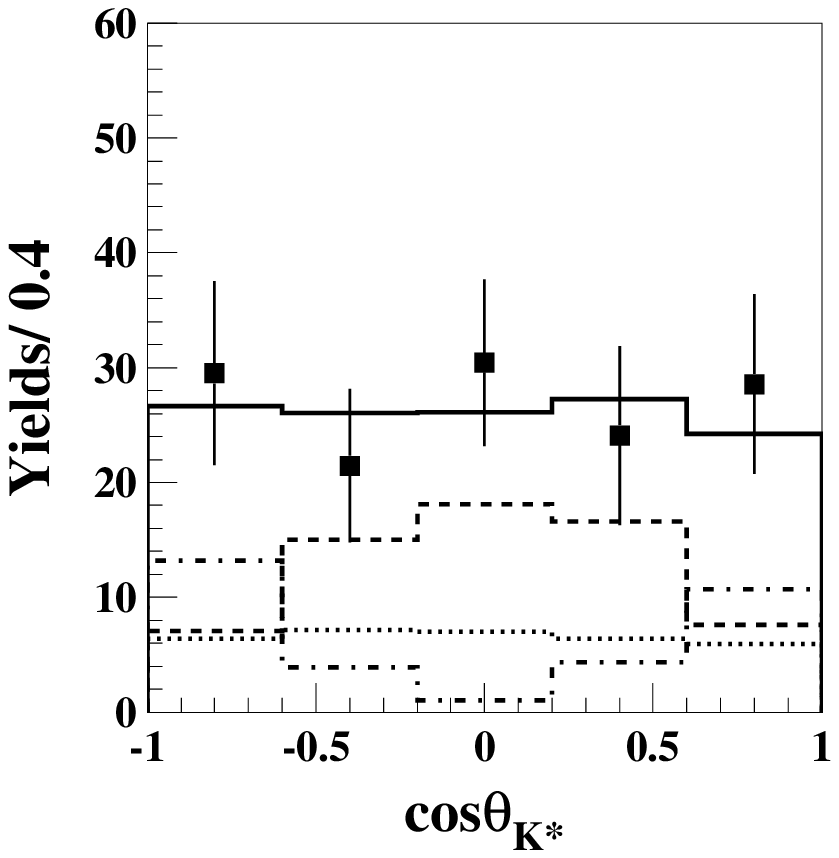}
    \caption{Fit to background-subtracted helicity distributions.
      The solid histograms show the results.
      The dot-dashed (dashed) histograms are the $A_0$
      ($A_\pm$) component of the fit; the dotted histograms are
      for non-resonant $\rho K\pi$. The low event yield near
      $\cos\theta_\rho=1$ is due to the
      $p_{\pi^0}>0.4{\,\rm GeV}/c$ requirement.}
    \label{fit_hel}
  \end{center}
\end{figure}

We use the $\rho^+ \to\pi^+\pi^0$ and $K^{*0}\to K^+\pi^-$
helicity-angle ($\theta_\rho$, $\theta_{K^*}$) distributions to
determine the relative strengths of $|A_{0}|^2$ and
$|A_{\pm}|^2$. Here $\theta_\rho$ ($\theta_{K^*}$) is the angle
between an axis anti-parallel to the $B$ flight direction and the
$\pi^+$ ($K^+$) flight direction in the $\rho$ ($K^*$) rest frame. 
For the longitudinal polarization case, the distribution is
proportional to $\cos^2 \theta_\rho \cos^2\theta_{K^*}$, and for the
transverse polarization case, it is proportional to $\sin^2\theta_\rho
\sin^2\theta_{K^*}$~\cite{angular}.
Figure~\ref{fit_hel} shows the signal yields obtained from $M_{\rm
  bc}$-$\Delta E$ fits in bins of the cosine of the helicity angle for
$\rho$ and $K^*$.
We perform a binned simultaneous $\chi^2$ fit to the 
$\rho$ and $K^*$ helicity-angle distributions.
The fit includes components for signal and non-resonant $\rho K\pi$. 
PDFs for signal $A_0$ and $A_\pm$ helicity states are determined from
the MC simulation.
The helicity-angle distribution for data in the high $M(K\pi)$
sideband region $1.1~{\rm GeV}/c^2<M(K\pi)<1.5$ GeV/$c^2$,
where $\rho K\pi$ events dominate, is consistent with a
$\cos^2\theta$-like $\cos \theta_\rho$ and a flat $\cos \theta_{K^*}$
distribution.
Thus, we assume an $S$-wave $K\pi$ system and model the non-resonant
$B \to \rho K \pi$ PDF based on the MC simulation.
The fraction of the non-resonant component is fixed at the values
obtained from the $K^*$ mass fit.
The two-parameter (normalizations for $A_0$ and $A_\pm$) fit result
deviates from 100\% longitudinal polarization with a  significance of
$4.9\,\sigma$ ($4.3\,\sigma$ including systematic uncertainties).
The significance is defined as $\sqrt{\chi_0^2-\chi_{\rm min}^2}$,
where $\chi_{\rm min}^2$ is the $\chi^2$ value at the best-fit and
$\chi_0^2$ is the value with the longitudinal polarization fraction
set to 100\%.

The largest uncertainties in the polarization measurement are due to
uncertainties in the non-resonant $\rho K\pi$ PDF, potential
scalar-pseudoscalar ($S$-$P$) interference, and the non-resonant
fraction. 
We assign a $^{+10.3}_{-0}$\% systematic error for the non-resonant
PDF. This uncertainty is estimated by adding a 1/3 flat component
to the $\rho$ helicity PDF for non-resonant $\rho K\pi$ in the
helicity fit. 
Interference of the longitudinal amplitude $A_0$ with the $S$-wave
$(K\pi)$ system introduces a term
with a $2 e^{i\Delta \phi}|A_{\rho K\pi}|\cos \theta_{K^*}$
dependence, where $\Delta \phi$ is the phase difference and $|A_{\rho
  K\pi}|$ is the amplitude of the $B\to \rho K\pi$ decay.
The $S$-$P$ wave interference disappears in the $\cos\theta_\rho$
distribution, which is integrated over $\cos\theta_{K^*}$; however it
remains in the $\cos\theta_{K^*}$ distribution.
We include an additional linear function for the interference term in
the $\cos\theta_{K^*}$ helicity, and redo the $\chi^2$ fit. 
The resulting small change in $f_L$, 0.5\%,
is assigned as the systematic uncertainty for the $S$-$P$ interference.
A $^{+4.0}_{-4.1}$\% systematic error is assigned for the uncertainty in
the fraction of non-resonant $\rho K\pi$, obtained by varying the
non-resonant fraction by $\pm 1\,\sigma$.
Adding the various systematic error contributions in quadrature, we
obtain the longitudinal polarization fraction in $B^+\to \rho^+
K^{*0}$ decays
$$f_L(B^+\to \rho^+ K^{*0})=0.43\pm0.11({\rm
  stat})^{+0.05}_{-0.02}({\rm syst}).$$

To calculate the $B^+\to \rho^+ K^{*0}$ branching fraction,
we use the $M(K\pi)$ invariant mass fit result and MC-determined
efficiencies weighted by the measured polarization components.
We consider systematic errors in the branching fraction that are
caused by uncertainties in the efficiencies of track finding, particle
identification, $\pi^0$ reconstruction, continuum suppression,
fitting, polarization fraction.
We assign an error of 1.1\% per track for the uncertainty in the
tracking efficiency. This uncertainty is obtained from a study of
partially reconstructed $D^*$ decays.
We also assign an uncertainty of 0.7\% per track on the particle
identification efficiency, based on a study of kinematically selected
$D^{*+}\to D^0\pi^+$, $D^0\to K^-\pi^+$ decay.
A 4.0\% systematic error for the uncertainty in the $\pi^0$ detection
efficiency is determined from data-MC comparisons of
$\eta\to\pi^0\pi^0\pi^0$ with $\eta \to\pi^+\pi^-\pi^0$ and $\eta \to
\gamma\gamma$.
A 4.5\% systematic error for continuum suppression is estimated from
studying the process $B^+ \to \overline{D}{}^0\pi^+$,
$\overline{D}{}^0\to K^+\pi^-\pi^0$.
A $-2.0\%/+1.7$\% systematic error associated with fits
is obtained by shifting each parameter by $\pm 1\,\sigma$.
A 6.7\% systematic error for the uncertainty in the $b\to c$
background PDF is obtained by changing the PDF parameterization. 
A $-4.2\%/+4.4$\% error due to the uncertainty in the fraction of
longitudinal polarization is obtained by varying $f_L$ by its errors. 
The uncertainty in non-resonant $K^* \pi \pi$ background gives a
contribution of $-2.2\%/+0\%$ in addition to $-3.0\%/+2.3\%$ error from
uncertainties in the background from other rare $B$ decays.
A 7.1\% error for possible bias in the $\chi^2$
fit~\cite{ref:chi2_bias} is obtained from a MC study.
A 1.1\% error for the uncertainty in the number of $B\overline B$ events
in the data sample is also included.
The quadratic sum of all of these errors is taken as the total
systematic error.
We obtain the branching fraction
$${\cal B}(B^+\to \rho^+K^{*0})=(8.9\pm 1.7({\rm
  stat})\pm 1.2 (\rm syst))\times 10^{-6}.$$

In summary, we have observed the $B^+ \to\rho^+ K^*$ decay with
a statistical significance of $5.3\,\sigma$.
We measure the branching fraction to be $(8.9\pm 1.7(\rm stat)\pm
1.2(\rm syst))\times 10^{-6}$.
We also perform a helicity analysis and find a substantial
transversely polarized fraction with a statistical significance of
$4.9\,\sigma$.
The longitudinal polarization fraction $f_L$ measured is similar to
the surprisingly low value found in $b \to ss\bar s$ decays
$B \to \phi K^*$.\\

We thank the KEKB group for the excellent operation of the
accelerator, the KEK cryogenics group for the efficient
operation of the solenoid, and the KEK computer group and
the NII for valuable computing and Super-SINET network
support.  We acknowledge support from MEXT and JSPS (Japan);
ARC and DEST (Australia); NSFC (contract No.~10175071,
China); DST (India); the BK21 program of MOEHRD and the CHEP
SRC program of KOSEF (Korea); KBN (contract No.~2P03B 01324,
Poland); MIST (Russia); MHEST (Slovenia);  SNSF (Switzerland); NSC and MOE
(Taiwan); and DOE (USA).


\begin{thebibliography}{99}
\bibitem{charge_conjugate} 
The inclusion of charge conjugate modes is implied unless stated otherwise.
\end{thebibliography}
\end{document}